\documentclass[amssymb,prl,aps,amsmath,showpacs,twocolumn]{revtex4-1}
\pdfoutput=1
\usepackage{graphicx}
\usepackage{amssymb}
\usepackage{amsmath}
\usepackage{bm}
\usepackage{epstopdf}

\begin{document}
\setcounter{page}{1}
\title[]{Observation of the universal jump across the Berezinskii-Kosterlitz-Thouless transition in two-dimensional Bose gases}

\author{Jiho \surname{Noh}}

\author{Jeongwon \surname{Lee}}

\author{Jongchul \surname{Mun}}
\email{jcmun@kriss.re.kr}

\affiliation{Korea Research Institute of Standards and Science, Daejeon 305-340, Korea}
\date[]{}

\begin{abstract}
The physics in two-dimensional (2D) systems is very different from what we observe in three-dimensional (3D) systems. Thermal fluctuations in 2D systems are enhanced\cite{PRL17Mermin}, and they prevent the conventional Bose-Einstein condensation (BEC) at non-zero temperatures by destroying the long-range order. However, a phase transition to a superfluid phase is still expected to occur in a 2D system along with an emergence of a quasi-long-range order, explained by the Berezinskii-Kosterlitz-Thouless (BKT) mechanism\cite{SPJETP34Berenzinskii,JPC6Kosterlitz}. Within the BKT mechanism, a universal jump of the superfluid density in a 2D Bosonic system was theoretically predicted by Nelson and Kosterlitz\cite{PRL39Nelson}, and was first observed in 2D \textsuperscript{4}He films by Bishop and Reppy\cite{PRL40Bishop}. Recent experiments in trapped ultracold 2D Bose gas systems have shown signatures of the BKT transition\cite{Nature441Hadzibabic,PRL99Krueger,NJP10Hadzibabic,PRL102Clade,PRL105Tung,Nature470Hung}, and its superfluidity\cite{NATPHYS8Desbuquois}. However, the universal jump in the superfluid density was not observed in these systems. Here we report the first observation of the universal jump in the superfluid density using an optically trapped ultracold 2D Bose gas. The measured superfluid phase space density at the BKT transition agrees well with the predicted value within our measurement uncertainty. Additionally, we measure the phase fluctuations in our density profiles to show that the BKT transition occurs first, followed by the BEC transition\cite{PRA77Simula}.
\end{abstract}
\maketitle

Many features of the BKT transition have been revealed in 2D Bose gas systems \cite{Nature441Hadzibabic,PRL99Krueger,NJP10Hadzibabic,PRL102Clade,PRL105Tung,Nature470Hung,PRA84Plisson}. In particular, there has been a controversy about whether the sudden increase of a low-momentum coherence peak could be observed, and be considered as a signature of the BKT transition\cite{PRL105Tung,PRA84Plisson}. However, none of the previous studies have observed the sudden universal jump of the superfluid density. In a uniform system, the BKT mechanism predicts the universal jump in the superfluid phase space density from $n_{s}\lambda^{2}=0$ to $n_{s}\lambda^{2}=4$ at the transition\cite{PRL39Nelson}, where $n_{s}$ is the superfluid density, $\lambda=(2\pi\hbar^{2}/mk_{B}T)^{1/2}$ is the thermal wavelength, $\hbar=h/2\pi$ is the reduced Planck constant, $m$ is the mass of the atom, $k_B$ is the Boltzmann constant, and $T$ is the temperature of the gas.

In this letter, we report the first observation of the universal jump of the superfluid density, measured at diverse BKT transition temperatures, by varying the temperature and the atom number of the 2D atomic cloud.
In addition, we determined the distribution of the \textit{in situ} phase fluctuation by measuring the density fluctuation after a ballistic expansion, which is a widely used method\cite{PRL87Dettmer,PRL102Clade,PRL109Choi}. The phase fluctuation increases at the transition, which coincides with the expected increase in the vortex population within the BKT-driven nature of the superfluid transition. Furthermore, we observed the suppression of the phase fluctuation at the superfluid core as we lowered the temperature, which indicates a crossover from the BKT-driven superfluid phase to the BEC-driven superfluid phase\cite{PRA77Simula}. A BEC-driven superfluid phase is expected to occur in a trapped 2D Bose gas at finite temperature\cite{PRA44Bagnato}, in which the phase fluctuation would be suppressed due to the extended phase coherence.

In our experiment, the preparation of a 2D Bose gas begins with a 3D \textsuperscript{87}Rb BEC ($N\sim 10^{7}$) confined in a magnetic trap, where the atoms are in their $|F=1,m_{F}=-1\rangle$ internal ground state\cite{JKPS61Noh}. Then, the atoms are transferred into a single optical dipole trap (ODT) with harmonic trap frequencies of $(\omega_x,\omega_y,\omega_z)=2\pi\times(3.5,4.2,335)$ Hz. We can assume that the atomic cloud is in the quasi-2D regime as $\hbar\omega_z\sim k_{B}T$, where the degree of freedom in $z$-axis is thermodynamically frozen, and most of the atoms are in the thermal ground state along the axis. Our characteristic dimensionless 2D interaction strength is given by $\tilde{g}=\sqrt{8\pi}(a_{s}/a_{z})\approx0.05$ where $a_{s}$ is the s-wave scattering length and $a_{z}$ is the harmonic-oscillator length along the tightly confined $z$-direction.

We performed an absorption imaging along the vertical, $z$-direction after 40 ms time-of-flight (TOF) expansion of the gas\cite{RNC34Hadzibabic}. An off-resonance imaging was used to avoid the multiple scattering effect\cite{PRA82Rath}, by detuning the probe beam $-1\Gamma$ away from the atomic resonance, where $\Gamma$ is the natural linewidth.
In the experiment, we varied the temperature and the atom number of the atomic cloud independently by adjusting the trap depth and the hold time, respectively, and observed the 2D density profiles (Fig. 1b, c). Below the critical phase space density, the density profiles of the atomic cloud show Gaussian distribution.

\begin{figure}
\includegraphics{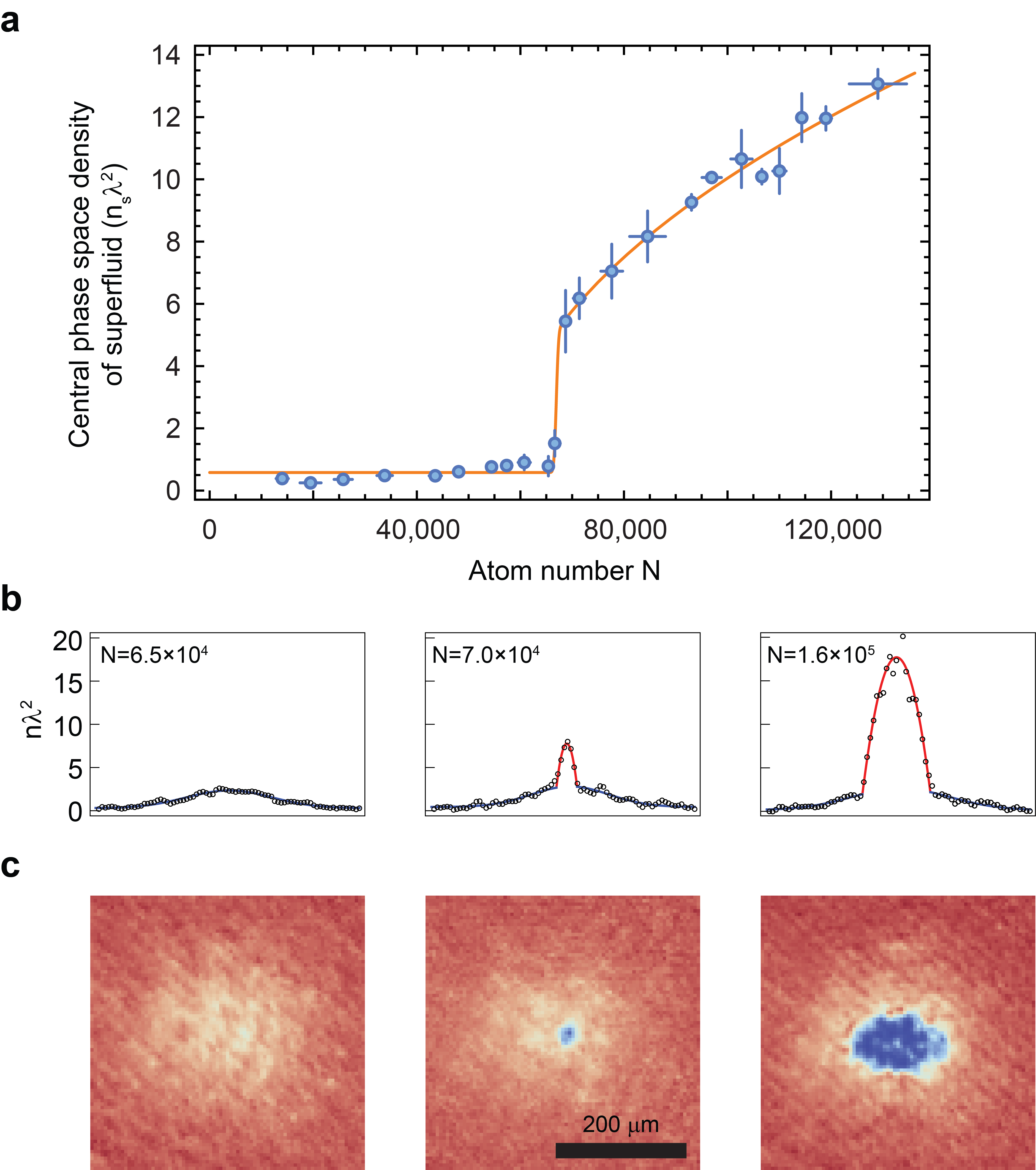}
\caption{\textbf{Superfluid jump varying the atom number.} \textbf{a,} Central phase space density of the superfluid ($n_{s}\lambda^{2}$) profiles as a function of the atom number. The temperature is kept constant at 27.4 nK. The solid line is a calculated fit to the superfluid central phase space density. \textbf{b,}  The cross-section of the two-dimensional profile along the $y$-axis of the atomic cloud centre, with the blue lines fit to the Gaussian distribution of the thermal component and the red lines fit to the Thomas-Fermi distribution of the superfluid component\cite{PRL100Holzmann}. Each plot is measured with a different atom number, $N$. \textbf{c,} Corresponding false-colour absorption images measured after 40 ms TOF. Error bars are standard deviations of the measurement.}
\label{fig1}
\end{figure}

At the critical point, a sharp central peak suddenly appears, and starting from this point, the density profiles show bimodality, which is composed of a Gaussian profile for the thermal gas component, and a narrow Thomas-Fermi (TF) profile for the superfluid component\cite{PRL100Holzmann,PRL99Krueger,PRL102Clade,RNC34Hadzibabic,PRA79Bisset}, of which the superfluidity was shown in a recent experimental work\cite{NATPHYS8Desbuquois}.
The temperature, $T$, was determined by fitting the thermal tail of the density profile with the mean-field Hartree-Fock (MFHF) theory\cite{NJP10Hadzibabic}.

Fig. 1a shows the central phase space density of the superfluid as a function of atom number at a constant temperature, $T=27.4\pm3.1$ nK. The central phase space density jumped abruptly from $n_{s}\lambda^{2}=0$ to $4.54\pm1.13$ near the critical atom number, $N\sim70000$. Such an abrupt feature of the jump in superfluid density has been calculated using Monte Carlo simulations\cite{PRA66Prokofev}, however, has been elusive in previous experiments\cite{Nature470Hung}, where only continuous change in quasi-condensate density was measured.

The jump in superfluid density was also observed by varying the temperature of the atomic cloud. Fig. 2 presents the central density of the superfluid as a function of temperature at a constant atom number, $N=1.7\times10^{5}$. In addition, a theoretical fit from the Monte Carlo calculation is plotted as a solid line in Fig. 2, in which the small change in trap frequency according to the changed trap depth is considered. Due to limitations such as the finite size effects\cite{RNC34Hadzibabic} and the shot to shot fluctuations of the temperature, the jump of the superfluid density could not be measured as a perfect step function at the boundary. These limitations round off the sudden jump of the superfluid phase space density to an error-function-like profile as shown in Fig. 2.
As the temperature decreases across the critical temperature, $T=75.2\pm4.7$ nK, the central density of the superfluid jumps from $n_{s}\lambda^{2}=0$ to $4.26\pm0.85$.
\begin{figure}
\includegraphics{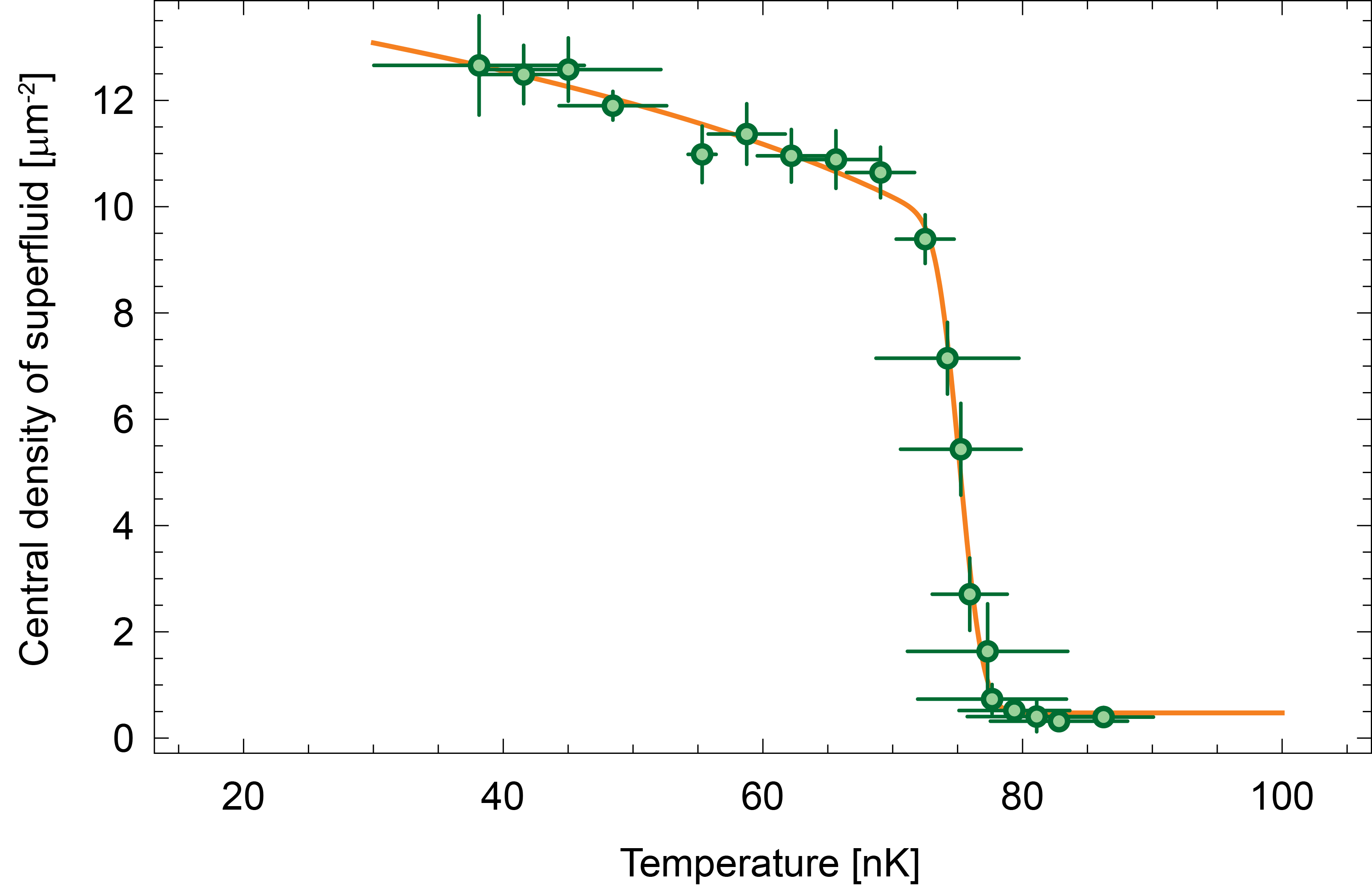}
\caption{\textbf{Superfluid jump varying the temperature.} Central superfluid density profiles for the superfluid transition as a function of the temperature. The total number of atoms is kept constant at $1.7\times10^{5}$. The dots indicate the central superfluid density. The solid line is a fit derived from the Monte Carlo calculations\cite{PRA66Prokofev} with considerations of changes in the trap frequencies as a function of temperature. Error bars are standard deviations of the measurement.}
\label{fig2}
\end{figure}

The universality of the jump in superfluid phase space density was observed by comparing our measurements at different transition temperatures (Fig. 3). Two more critical phase space densities were measured at 34.4 and 83.2 nK by varying the atom number. Combining our results, the central superfluid phase space density at the BKT transition was measured to be $n_{s}\lambda^{2}=4.23\pm0.47$, which agrees with the predicted value of $n_{s}\lambda^{2}=4$ within our measurement error. Such a universality strongly supports the BKT mechanism of our superfluid transition rather than the BEC-driven superfluid transition\cite{PRA44Bagnato}, where no abrupt jump is expected.
In addition, we note that our measurement precision was mainly limited by the atom number calibration error. More precise determination of the critical value would be possible by improving the imaging technique.
\begin{figure}
\includegraphics{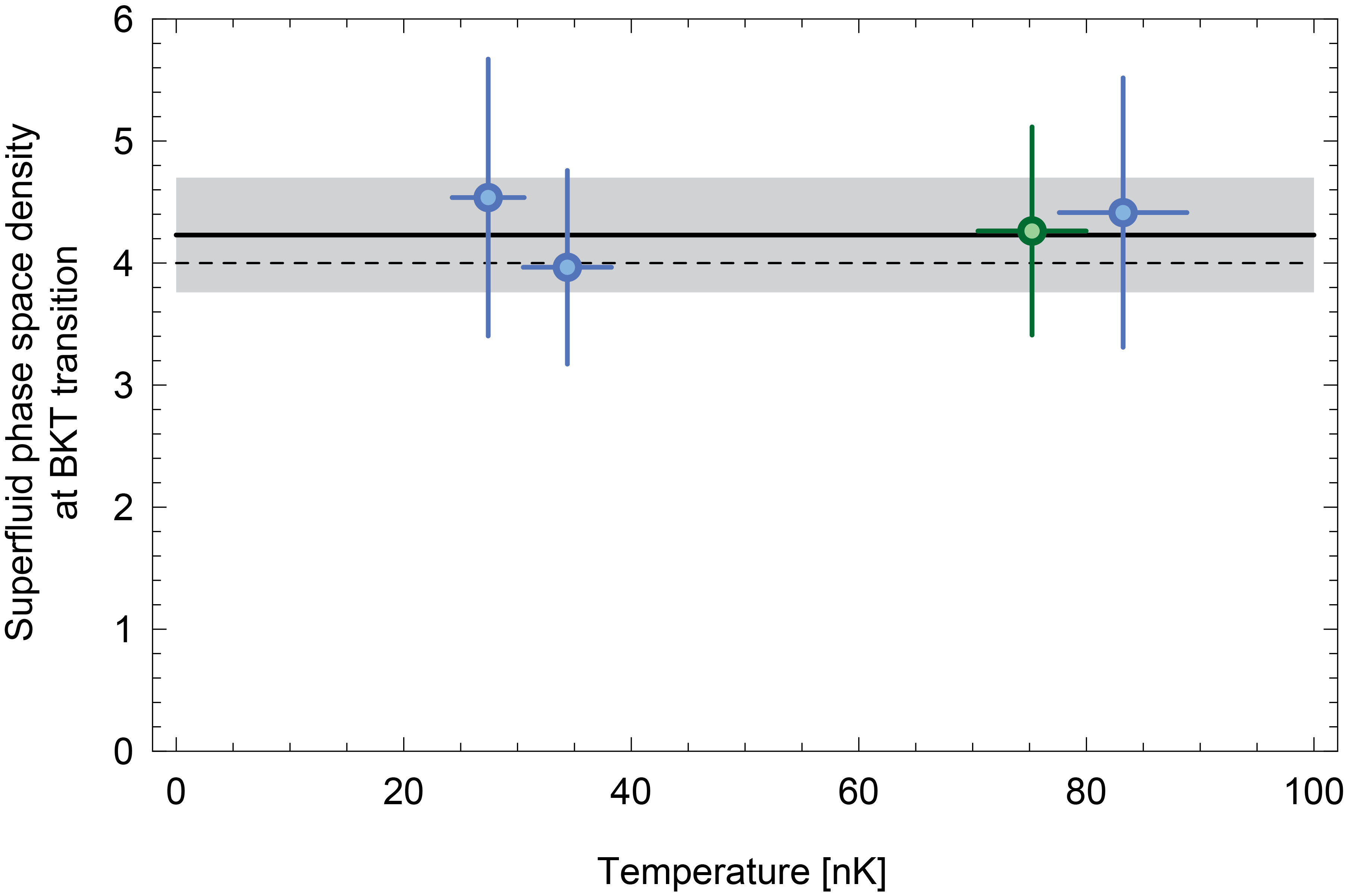}
\caption{\textbf{Universality of the jump in superfluid density.} Critical superfluid phase space densities measured at different transition temperatures. Blue dots and the green dot represent the measured critical superfluid phase space densities from experiments varying the atom number and the temperature, respectively. The solid line indicates the measured mean critical superfluid phase space density, $n_{s}\lambda^{2}=4.23\pm0.47$, and the shaded area marks the standard deviation of the measurements. The dashed line indicates the predicted critical value of $n_{s}\lambda^{2}=4$. Error bars are standard deviations of the measurement.}
\end{figure}

As another evidence of the BKT transition, we observed an increase in phase fluctuation at the transition, which is intrinsically related with the emergence of vortices and antivortices in a 2D Bose gas. It is theoretically predicted that the atomic gas in a superfluid phase can be divided into two regions: a region where a sudden increase in the probability of finding vortices and antivortices occurs (BKT-driven superfluid), and a vortex-free region at the core of the superfluid region (BEC-driven superfluid)\cite{PRL96Simula,PRA77Simula}. As in Fig. 1c, we observed the emergence of vortices along the boundary of the superfluid component at the atomic cloud with a large atom number. In a 2D Bose gas, the \textit{in situ} vortex core radius is set by the healing length, $\xi=1/\sqrt{\tilde{g}n}$, where $n$ is the density of atomic gas, and is theoretically expected to be observable within the imaging resolution of our system\cite{PRA61Dalfovo}. However, the phase fluctuations such as the inherent thermal fluctuations in the 2D system add complicated density ripples during the free expansion\cite{PRL109Choi}. As a consequence, the visibility of vortices in the TOF images is suppressed, which makes the vortex population measurement difficult.

For an alternative method, we measured the density fluctuations across the atomic gas after 30ms TOF, from which we deduced the \textit{in situ} phase fluctuations\cite{PRL87Dettmer}. We measured the standard deviation of the density distribution ($\Delta N$) along the ellipsoidal strips of equal width, and normalised it with the mean density ($\langle N\rangle$) along the strip\cite{PRL87Dettmer}.
The density fluctuation was detected using the CCD camera with high quantum efficiency ($\sim97\%$) to minimise detector shot noise. Furthermore, in order to rule out the effect coming from the photon shot noise, we subtracted out its effect from the variance in the measured optical density ($\Delta OD$)\cite{PRL105Sanner}.

Fig. 4a shows the density fluctuation distribution at three different temperatures.
At high temperatures without the superfluid fraction, there is no discernible density fluctuation. As we lower the temperature, the superfluid domain starts to emerge from the centre, along with the increase in fluctuation at the transition region. These effects are in good agreement with the predicted emergences of vortices and antivortices indicating the BKT-driven superfluid transition\cite{PRA77Simula}. As we decreased the temperature even further, the superfluid domain becomes larger and the phase fluctuation becomes suppressed deep inside the superfluid region (BEC-driven superfluid) while both peak position and inner radius of the BKT-driven superfluid move away from the fluctuation suppressed core (Fig. 4b). Such an increase of the inner radius of the BKT-driven superfluid shows the crossover from the BKT-driven superfluid phase to the BEC-driven superfluid phase at the core of the atomic cloud.

We have presented the first direct observation of the universal jump in superfluid phase space density by using an optically trapped ultracold 2D Bose gas. Along with the observation of increased phase fluctuation at the transition region, our observation of the universal jump supports the BKT mechanism. In addition, suppression of the density fluctuations at the core of superfluid region was observed, which indicates a crossover from the BKT-driven superfluid to the BEC-driven superfluid.
The superfluid density, which we measured using bimodal profile\cite{PRL100Holzmann,PRL102Clade,PRA79Bisset}, could be measured with more direct methods such as proposed in ref. 27.
Further investigation of the transition point with various interaction strengths, $\tilde{g}$, would provide more insights into the universality of superfluid jump, and phase diagram (normal, BKT, BEC) of 2D Bose gas on both temperature and interaction planes\cite{Nature441Esslinger}.

\begin{figure}[b]
\includegraphics{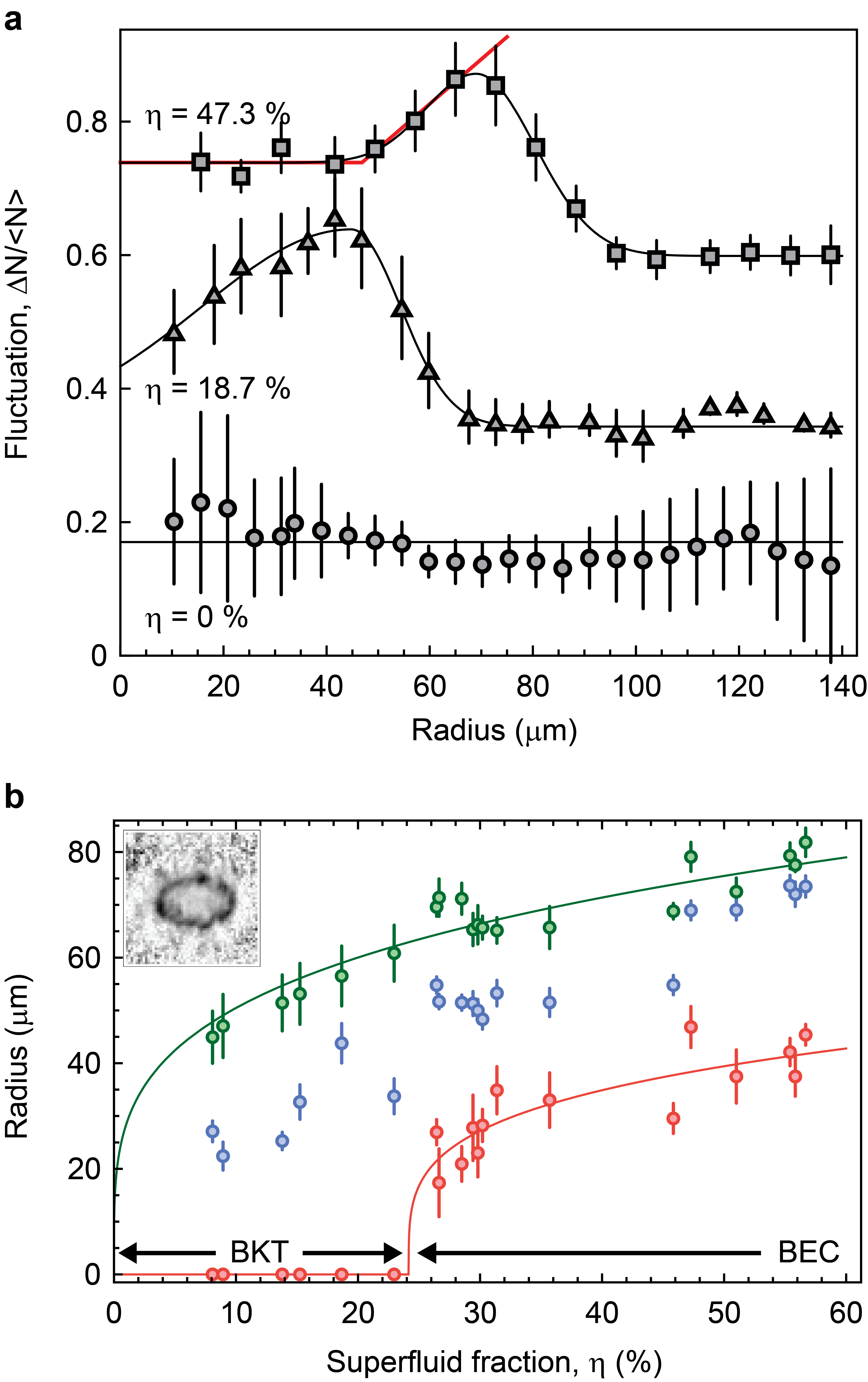}
\caption{\textbf{The distribution of atomic density fluctuation after ballistic expansion and the measurement of the BKT-driven superfluid.} \textbf{a,} Normalised atomic density fluctuation ($\Delta N/\langle N \rangle$) as a function of the radial distance from the centre of the atomic cloud, measured at representative superfluid fractions: 0\% (circle), 18.7\% (triangle), and 47.3\% (square). Each distribution is displayed with an offset for clarity. Black solid lines are fits for finding the peak of the distribution, and red solid lines for determining the boundary between the BKT-driven and the BEC-driven superfluid domains. \textbf{b,} Blue dots and red dots indicate the peak positions and inner radii of the BKT-driven superfluid, respectively, as a function of the condensate fraction, $\eta$. Green dots indicate the Thomas-Fermi radii, $R_{TF}$, of the superfluid components. Solid lines are guide to the eye. Inset shows the representative plot of a density fluctuation distribution derived from the bimodal density distribution such as the third plot of Fig. 1c. It clearly shows three different domains in terms of phase fluctuations: fluctuation suppressed core, fluctuating superfluid region and thermal background. Error bars are standard deviations of the measurement.}
\end{figure}

This work is supported by KRISS creative research initiative.

\end{document}